\documentclass[final,5p,twocolumn]{elsarticle}

\usepackage{graphicx}          
\usepackage[dvips]{epsfig}    

\usepackage{xcolor}

\usepackage{amssymb}
\usepackage{amsthm}

\usepackage{amsmath}          

\usepackage{algorithmic}

\begin{document}

\begin{frontmatter}

\title{Optimal nonlinear damping control of second-order systems} 

\author{Michael Ruderman}
\ead{michael.ruderman@uia.no}


\address{University of Agder, 4604-Norway}

\begin{abstract}                          
Novel nonlinear damping control is proposed for the second-order
systems. The proportional output feedback is combined with the
damping term which is quadratic to the output derivative and
inverse to the set-point distance. The global stability, passivity
property, and convergence time and accuracy are demonstrated. Also
the control saturation case is explicitly analyzed. The suggested
nonlinear damping is denoted as optimal since requiring no design
additional parameters and ensuring a fast convergence, without
transient overshoots for a non-saturated and one transient
overshoot for a saturated control configuration.
\end{abstract}


\end{frontmatter}

\newtheorem{thm}{Theorem}
\newtheorem{lem}[thm]{Lemma}
\newtheorem{clr}{Corollary}
\newdefinition{rmk}{Remark}
\newproof{pf}{Proof}

\section{Introduction}
\label{sec:0}

For the second-order systems, it is understood that a linear
feedback control \cite{franklin2015} inherently poses certain
limits in terms of possibility to shape the transient response,
exponential convergence of the state trajectories and, as
implication, steady-state accuracy of the controlled output of
interest. Worth to recall is that the input-output second-order
systems encompass a vast number of practical applications. Input
voltage to output speed in motors, transfer characteristics of
different-type RLC circuits, pressure-flow dynamics in the fluid
transport systems and, finally, motion dynamics of rigid-body
systems, in general sense, can be noted as motivating examples for
that.

For linear control systems, an assignment of optimal damping, so
as to shape the desired dynamic response, is straightforward
through for instance pole placement, cf. e.g. with
\cite{franklin2015}. Also when allowing for a system damping to be
switched once as a function of the system state, an optimal
damping ratio for linear second-order systems has been proposed in
the past \cite{shahruz1992}. On the other hand, nonlinear control
methodology addressed, since long, the problem of an efficient
feedback shaping, while the complexity of associated analysis and
control synthesis, availability of the system states, control
specification, and type of the system perturbations led to quite
different design concepts. Among the well-established are the
sliding mode control \cite{shtessel2014sliding}, Lypunov redesign
\cite{parks1966liapunov}, backstepping \cite{kokotovic1992joy},
and passivity based control \cite{ortega2013passivity}. For more
details and well-known basics we also refer to seminal literature
e.g. \cite{SlotLi90,khalil2002}. Some former examples of the
nonlinear feedback stabilization and associated nonlinear damping
can be found in e.g.
\cite{kanellakopoulos1992toolkit,teel1992global} to mention few
here. A comparative evaluation of different controllers,
benchmarked on a most simple second-order plant of double
integrator, can also be found in \cite{rao2001naive}.

The need to incorporate nonlinear damping in feedback of the
second-order systems, especially for improving the stabilizing and
convergence properties, has been (empirically) recognized in
already former studies in robotics, thus resulting in e.g.
nonlinear proportional-derivative controls
\cite{HolerPDnonlin,kelly1996}. While the stability proof has been
provided for several ad-hoc nonlinear damping strategies, no
optimal convergence and trajectories shaping have been so far
elaborated. Here it is also worth to side note that the
convergence properties are strongly related to homogeneity of the
corresponding dynamics vector-field and, as implication, of the
feedback map to be determined, in other words to be assigned. For
overview on the use of homogeneity for synthesis in, e.g. sliding
mode control, we exemplary refer to e.g.
\cite{bernuau2014homogeneity}. As another approach, to feedback
control problems, it appears that to enter energy into a system,
through potential field of the output feedback, is more
straightforward than to control its dissipation. The latter should
occur in a peculiar way, thus ensuring the desired convergence to
an equilibrium. For energy shaping in the feedback regulated
Euler-Lagrange systems we refer to e.g. \cite{loria1997global} and
some basic literature \cite{ortega2013passivity}.

In this paper, we propose a novel nonlinear damping control of the
second-order systems in combination with the linear output
feedback. Using the fact of conservative energies exchange in an
undamped (oscillatory) second-order system, the dissipated energy
is shaped in an optimal way with respect to the convergence to
zero equilibrium and no transient overshoot independent of the
initial state. That way assigned nonlinear damping is quadratic to
the output derivative and inverse to the set-point distance, while
no free design parameters for the damping term are required. The
proposed control is generic and globally asymptotically stable. It
also allows for control saturation, that is relevant for
applications. The principle analysis of the control behavior,
provided below, is focusing on the unperturbed second-order
dynamics only. In that was the further aspects of sensitivity and
robustness are subject to the future works.

\section{Second order system with state-feedback}
\label{sec:1}

Throughout the paper we will deal with the feedback controlled
second-order systems
\begin{eqnarray}
\label{eq:1}
  \dot{x}_1 &=& x_2, \\
  \dot{x}_2 &=& -k x_1 -D,
\label{eq:2}
\end{eqnarray}
where $x_1$ and $x_2$ are the available state variables, $k>0$ is
the proportional feedback gain, and $D$ is the control damping of
interest, correspondingly to be shaped. Obviously, the system
\eqref{eq:1}, \eqref{eq:2} is a classical double-integrator
dynamics, for which a vast number of application examples can be
found e.g. in electrical and mechanical systems and combinations
of those.

\subsection{Optimal linear damping} \label{sec:1:sub:1}

Using the linear state-feedback damping, the system \eqref{eq:1},
\eqref{eq:2} can be written in a standard state-space form
\begin{equation}\label{eq:3}
[\dot{x}_1,\dot{x}_2]^T =A \cdot [x_1,x_2]^T =
\left[%
\begin{array}{cc}
  0 & 1 \\
  -k & -d \\
\end{array}%
\right]
\cdot[x_1,x_2]^T,
\end{equation}
where the system matrix $A$ is Hurwitz, for positive damping
coefficients $d>0$, and is already in the controllable canonical
form. It is worth recalling that the state-feedback controlled
system \eqref{eq:3} is equivalent to the proportional derivative
(PD) controller for which an appropriate choice of the feedback
gains allow for arbitrary shaping the closed-loop response, either
in time $t$- or in Laplace $s$-domain. Assuming that $k$ is given
(by some control specification) and requiring the control response
has no transient oscillations or overshoot, meaning the real poles
only, one can assign the linear damping term by solving
\begin{equation}\label{eq:4}
s^2+ds+k=(s+\lambda)^2
\end{equation}
with respect to $d$. Here the real double-pole at $-\lambda$
determines the optimal linear damping, usually noted as critical
damping, since for $d>2\lambda$ the system behaves as overdamped,
while for $d<2\lambda$ the system becomes transient oscillating.
For any non-zero initial conditions $[x_1,x_2]^T(0)\equiv[x_1^0,
x_2^0]^T\neq0$, which can be seen as a set-value control problem,
the trajectories are given by
\begin{equation}\label{eq:5}
[x_1,x_2]^T(t)= \exp(At) \, [x_1^0, x_2^0]^T.
\end{equation}
It is obvious that the unperturbed matrix differential equation
\eqref{eq:3}, with two stable real poles, has an exponential
convergence property, meaning
\begin{equation}\label{eq:6}
\|x_1(t),x_2(t)\| \leq \beta \exp(-\gamma t)
\end{equation}
for some $\beta, \gamma >0$ constants. From the output control
viewpoint that means $x_1 \rightarrow 0 $ for $t \rightarrow
\infty$.

\section{Main results}
\label{sec:2}

\subsection{Optimal nonlinear damping} \label{sec:2:sub:1}

\begin{figure}[!h]
\centering
\includegraphics[width=0.98\columnwidth]{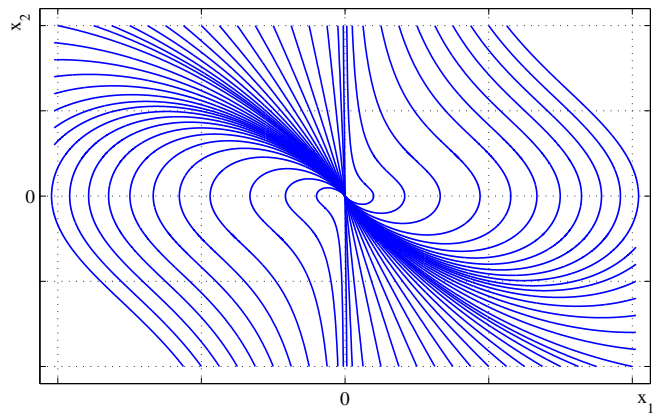}
\caption{Phase portrait of the control system \eqref{eq:7a},
\eqref{eq:7b}.} \label{fig:1}
\end{figure}

The proposed nonlinear damping endows the system \eqref{eq:1},
\eqref{eq:2} to be
\begin{eqnarray}
\label{eq:7a}
  \dot{x}_1 &=&x_2,  \\
  \dot{x}_2 &=&-kx_1-x_2^2|x_1|^{-1} \mathrm{sign}(x_2) \label{eq:7b}.
\end{eqnarray}
The single control parameter remains the given output feedback
gain, while the quadratic damping term yields optimal for all
$k>0$ values. The solution of \eqref{eq:7b} is non-singular except
in $x_1=0$, while the unique equilibrium $(x_1,x_2)=0$ is globally
attractive as will be shown below in section \ref{sec:2:sub:2}.
The phase portrait of the system \eqref{eq:7a}, \eqref{eq:7b} is
shown in Fig. \ref{fig:1}.
One can recognize that the damping rate, and the required control
effort, which is $\sim \dot{x}_2$, notably increases in vicinity
of $x_1=0$ for $|x_2| \gg 0$. At the same time, the non-singular
solution provides the global convergence to origin within the II
and IV quadrants without $x_1$ zero crossing, thus without
transient overshoot of the control response. For showing this,
consider the region of attraction in vicinity of the origin. For
the steady-state one obtains
\begin{equation}\label{eq:8}
\mathbf{0}=\left[%
\begin{array}{cc}
  0  & 1 \\
  -k & -|x_2||x_1|^{-1} \\
\end{array}%
\right]\cdot[x_1,x_2]^T,
\end{equation}
which results in
\begin{equation}\label{eq:9}
k|x_1|x_1=-|x_2|x_2.
\end{equation}
This can be seen as a trajectories' attractor in vicinity of zero
equilibrium. Rewriting \eqref{eq:9} as
\begin{equation}\label{eq:10}
kx_1^2  \,\mathrm{sign}(x_1)=-x_2^2 \, \mathrm{sign}(x_2)
\end{equation}
and allowing for the real solution only, results in
\begin{equation}\label{eq:11}
x_2+ \sqrt{k} x_1=0,
\end{equation}
which is a slope along which the trajectories converge to zero in
vicinity of origin, that without crossing the $x_2$-axis.

\subsection{Global stability} \label{sec:2:sub:2}

Assume the following Lyapunov function candidate
\begin{equation}\label{eq:12}
V=\frac{1}{2}x_2^2+k\frac{1}{2}x_1^2,
\end{equation}
which is positive definite for all $(x_1,x_2) \neq 0$ and also
radially unbounded, i.e. $V(x_1x_2) \rightarrow \infty$ as
$\|x_1,x_2\| \rightarrow \infty$. Taking the time derivative and
substituting dynamics of the states, i.e. \eqref{eq:7a} and
\eqref{eq:7b}, results in
\begin{equation}\label{eq:13}
\dot{V}=-x_2^3|x_1|^{-1} \mathrm{sign}(x_2)\leq 0,
\end{equation}
which implies the origin is globally stable. Since the
trajectories do not remain on the $x_1$-axis when $x_2=0$, due to
non-zero vector field cf. \eqref{eq:7b}, and proceed towards
origin, cf. Fig. \ref{fig:1}, the asymptotic stability of origin
can be concluded despite $\dot{V}=0$ for $x_2=0$, $x_1\neq 0$.
This excludes an appearance of invariant sets and ensures $(0,0)$
is the single asymptotically stable equilibrium.

\subsection{Closed-loop passivity} \label{sec:2:sub:3}

\begin{figure}[!h]
\centering
\includegraphics[width=0.98\columnwidth]{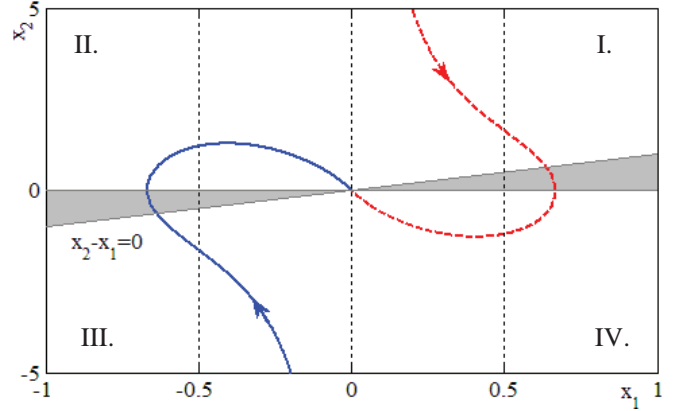}
\caption{Regions of system passivity in the phase plane.}
\label{fig:2}
\end{figure}

For analyzing damping properties of the control system
\eqref{eq:7a}, \eqref{eq:7b} we are to demonstrate the passivity
of the closed-loop dynamics
\begin{equation}\label{eq:14}
\dot{x}_2+kx_1=-x_2^2|x_1|^{-1} \mathrm{sign}(x_2).
\end{equation}
Here the left-hand side can be seen as a conservative
(oscillatory) system part, in other words plant, and the
right-hand side of \eqref{eq:14} as a stabilizing control input
$u$ which provides the closed-loop system with a required damping.
Recall that for a system with output $y$ to be passive, the
input-output port power should be greater than or equal to the
rate of energy stored in the system self, i.e. $uy\geq \dot{V}$.
Here the same energy function as the Lyapunov function candidate
\eqref{eq:12}, which is the system's Hamiltonian, is assumed while
$x_1$ is the controlled system output of interest. The above power
inequality (for system passivity) yields
\begin{equation}\label{eq:15}
-x_2^2 \, \mathrm{sign}(x_2)\mathrm{sign}(x_1)\geq -x_2^2 \,
|x_2||x_1|^{-1},
\end{equation}
which results in the following passivity condition
\begin{equation}\label{eq:16}
\frac{|x_2|}{|x_1|}\geq \mathrm{sign}(x_2) \, \mathrm{sign}(x_1)
\end{equation}
for the state-space. Based on that it is evident that the system
is always passive in the II and IV quadrants of the phase plane,
see Fig. \ref{fig:2}. Otherwise, the system becomes transiently
non-passive for $x_2-x_1 < 0$ in the I quadrant and for $x_2-x_1
> 0$ in the III quadrant (gray-shadowed in Fig. \ref{fig:2}).
In those non-passive segments, the level of energy stored in the
system increases, this way also ensuring the state trajectories
always cross $x_1$-axis and do not remain at $x_2=0$. Following to
that, the trajectories always change, upon the velocity zero
crossing, to the passive segments of II or IV quadrant, which both
act as a control attractor to the globally stable origin.

\subsection{Convergence time} \label{sec:2:sub:4}

\begin{figure}[!h]
\centering
\includegraphics[width=0.98\columnwidth]{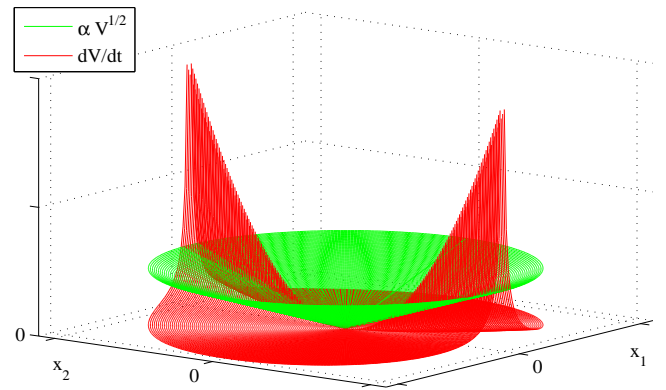}
\caption{Surface of time derivative of the Lyapunov function
$\dot{V}$ (in red) and square root of the Lyapunov function
$\alpha V^{1/2}$ (in green).} \label{fig:3}
\end{figure}

The asymptotic convergence of the state solutions is ensured by
$\dot{V}<0$. On the other hand, in order to ensure a finite-time
convergence, one has to show that
\begin{equation}\label{eq:17}
\dot{V}+\alpha V^{\frac{1}{2}}\leq 0
\end{equation}
for some positive time constant $\alpha> 0$. If inequality
\eqref{eq:17} holds, the finite convergence time $t_c$ is bounded
by
\begin{equation}\label{eq:18}
t_c \leq  2V^{\frac{1}{2}}(0) \alpha^{-1}.
\end{equation}
Substituting the Lyapunov function candidate \eqref{eq:12} and its
time derivative \eqref{eq:13} into \eqref{eq:17} results in
\begin{equation}\label{eq:19}
\frac{x_2^3}{|x_1|} \mathrm{sign}(x_2)\geq \alpha
\frac{\sqrt{2}}{2} \sqrt{x_2^2+kx_1^2}.
\end{equation}
An explanatory graphical interpretation of inequality
\eqref{eq:19} is shown in Fig. \ref{fig:3} by two surfaces, one of
the energy level and another of its time derivative.
One can recognize that the finite-time convergence can be ensured
in vicinity of $x_1=0$ and that until certain neighborhood to the
origin only (cf. both both red horns above the green cone in Fig.
\ref{fig:3}). Outside of those regions the inequality
\eqref{eq:19} becomes violated, cf. Fig. \ref{fig:3}, and the
control system \eqref{eq:7a}, \eqref{eq:7b} features the
asymptotic convergence. Here it is worth emphasizing that, from
the applications' viewpoint, such partial finite-time convergence
can be desired and sufficient, since the convergence to absolute
zero is inherently restricted by some finite resolution of sensors
used in the feedback control.

\subsection{Control with saturation}
\label{sec:2:sub:5}

From the applications' viewpoint, where frequently the input
limitations have to be taken into account, the control value $v$
with saturation is essential. That means it is to show whether the
proposed nonlinear damping control system \eqref{eq:7a},
\eqref{eq:7b} remains further on performing and, above all,
globally stable when the control input
\begin{equation}\label{eq:20}
-S \; \leq \; v = -kx_1-x_2^2|x_1|^{-1} \mathrm{sign}(x_2) \; \leq
\; +S
\end{equation}
is limited in the amplitude by some positive $S$. The latter
constitutes an inherent control system constraint, correspondingly
a given fixed parameter. In the saturated control mode, the system
\eqref{eq:7a}, \eqref{eq:7b} evolves to
\begin{eqnarray}
\label{eq:21}
  \dot{x}_1 &=& x_2, \\
  \dot{x}_2 &=& S \, \mathrm{sign}(v),
\label{eq:22}
\end{eqnarray}
and it has to be to proven whether the control value returns to
$|v(t)| < S$ after transients and, therefore, to the nominal
dynamic behavior independent of the initial conditions. In this
case one one needs to demonstrate that
\begin{equation}\label{eq:23}
\Bigl| -kx_1-x_2^2|x_1|^{-1} \mathrm{sign}(x_2) \Bigr| < S
\end{equation}
can be achieved and will hold for some finite time $t > 0$ for any
initial state $[x_1,x_2]^T(0) = [X_1, X_2]^T$. Due to symmetry of
the control system and without loss of generality we focus, in the
following, on the positive saturation only, while the respective
developments for a negative saturation are equivalent when turning
the sign and flipping the following inequality. The positive
control saturation requires to prove
\begin{equation}\label{eq:24}
-kx_1-x_2^2|x_1|^{-1} \mathrm{sign}(x_2) < S,
\end{equation}
while the saturated control action $\dot{x}_2 = S$ yields an
explicit solution of the state trajectories
\begin{eqnarray} \label{eq:25}
  x_2(t) &=& X_2 + St, \\
  x_1(t) &=& X_1 + \frac{1}{2} \, S t^2.
\label{eq:26}
\end{eqnarray}
Substituting \eqref{eq:25}, \eqref{eq:26} into \eqref{eq:24}
results in
\begin{equation}\label{eq:27}
-k \Bigl(X_1 + \frac{1}{2} \, S t^2 \Bigr) - \frac{\bigl(X_2+S
t\bigr)^2 \, \mathrm{sign}\bigl(X_2+ S t\bigr)} {\bigl| X_1 +
\frac{1}{2} S t^2 \bigr|} < S.
\end{equation}
While the second left-hand side term of \eqref{eq:27} remains
always positive for $t > -X_2 S^{-1}$, the brackets of the first
term remains also always positive for $t^2 > -2 X_1 S^{-1}$. This
implies that there is a $\tau > 0$ so that the condition
\eqref{eq:27} holds for all $t > \tau$. This proves the
closed-loop control system \eqref{eq:21}, \eqref{eq:22} always
returns to a non saturated control mode, i.e. \eqref{eq:7a},
\eqref{eq:7b}, at some $0 < t = \tau < \infty$, and that for all
admissible $\{X_1, X_2\}$ initial states and admissible control
parameters $S,k > 0$.

\section{Comparative numerical study}
\label{sec:3}

\begin{figure}[!h]
\centering
\includegraphics[width=0.98\columnwidth]{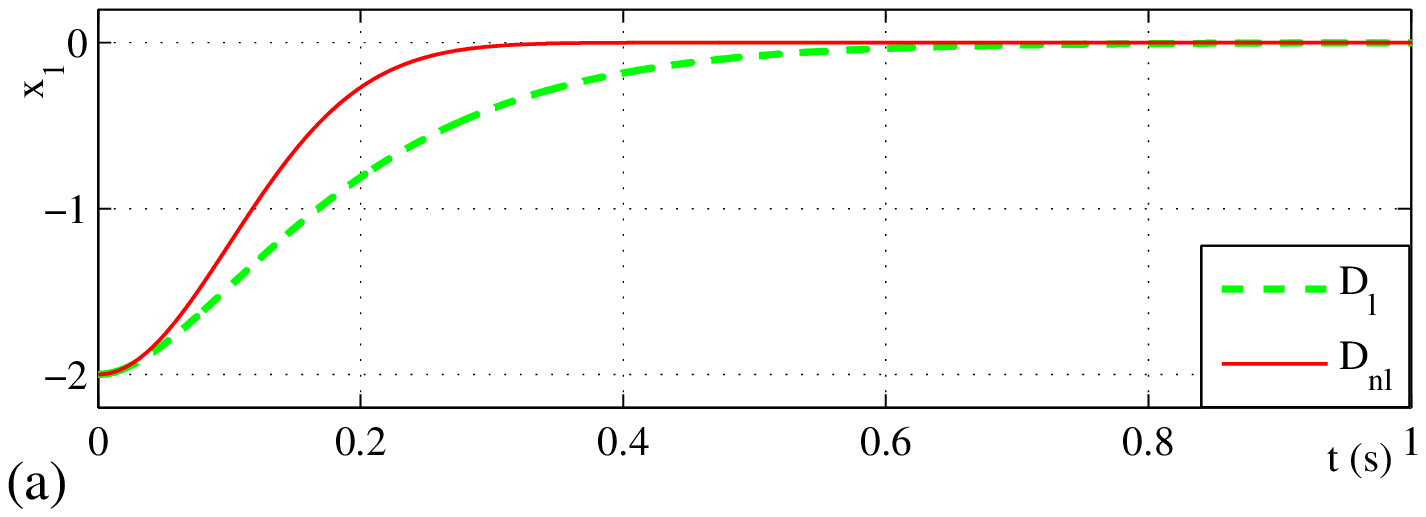}
\includegraphics[width=0.98\columnwidth]{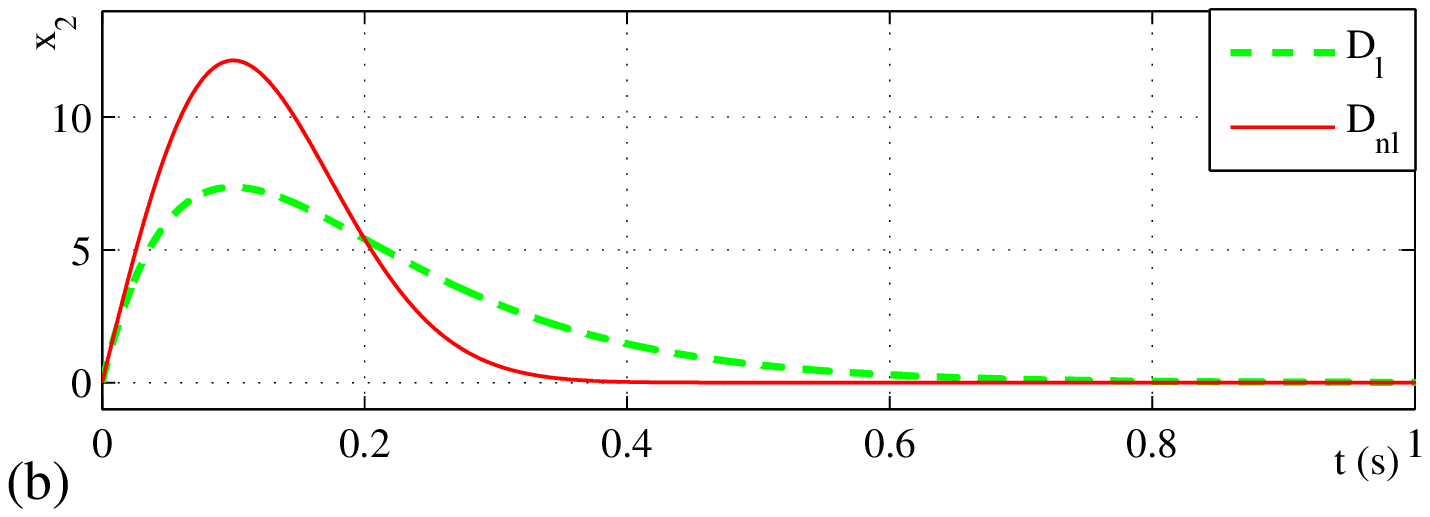}
\caption{Convergence of linear and nonlinear damping control.}
\label{fig:4}
\end{figure}

Two feedback control systems described by \eqref{eq:1},
\eqref{eq:2} are compared: one with the linear damping
$D_{l}=dx_2$ and one with the proposed nonlinear damping
$D_{nl}=x_2^2 |x_1|^{-1} \mathrm{sign}(x_2)$. The convergence of
the state trajectories is comparatively shown in Fig. \ref{fig:4}
for the initial values $[x_1^0,x_2^0]=(1,0)$ and output feedback
gain assigned to $k=100$. The optimal (critical) linear damping
factor, cf. \eqref{eq:4}, is $d=20$.

\begin{figure}[!h]
\centering
\includegraphics[width=0.98\columnwidth]{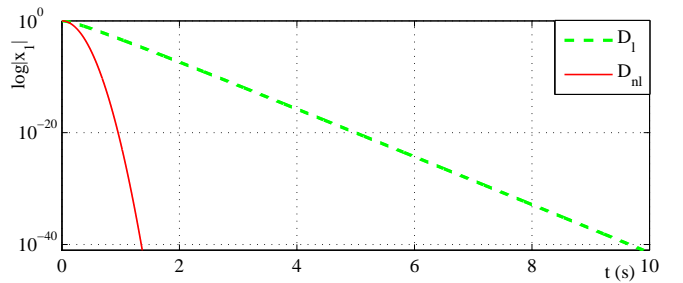}
\caption{Logarithmic convergence of the output magnitude value.}
\label{fig:5}
\end{figure}

As next, the convergence of the controlled output (absolute value)
is shown logarithmically in Fig. \ref{fig:5} for both the linear
and nonlinear damping. It can be seen that the control with
nonlinear damping reaches much faster, in fact quadratically on
the logarithmic scale, some low bound of the steady-state
accuracy. Different, the control with linear damping converges
linearly on the logarithmic scale.

\begin{figure}[!h]
\centering
\includegraphics[width=0.98\columnwidth]{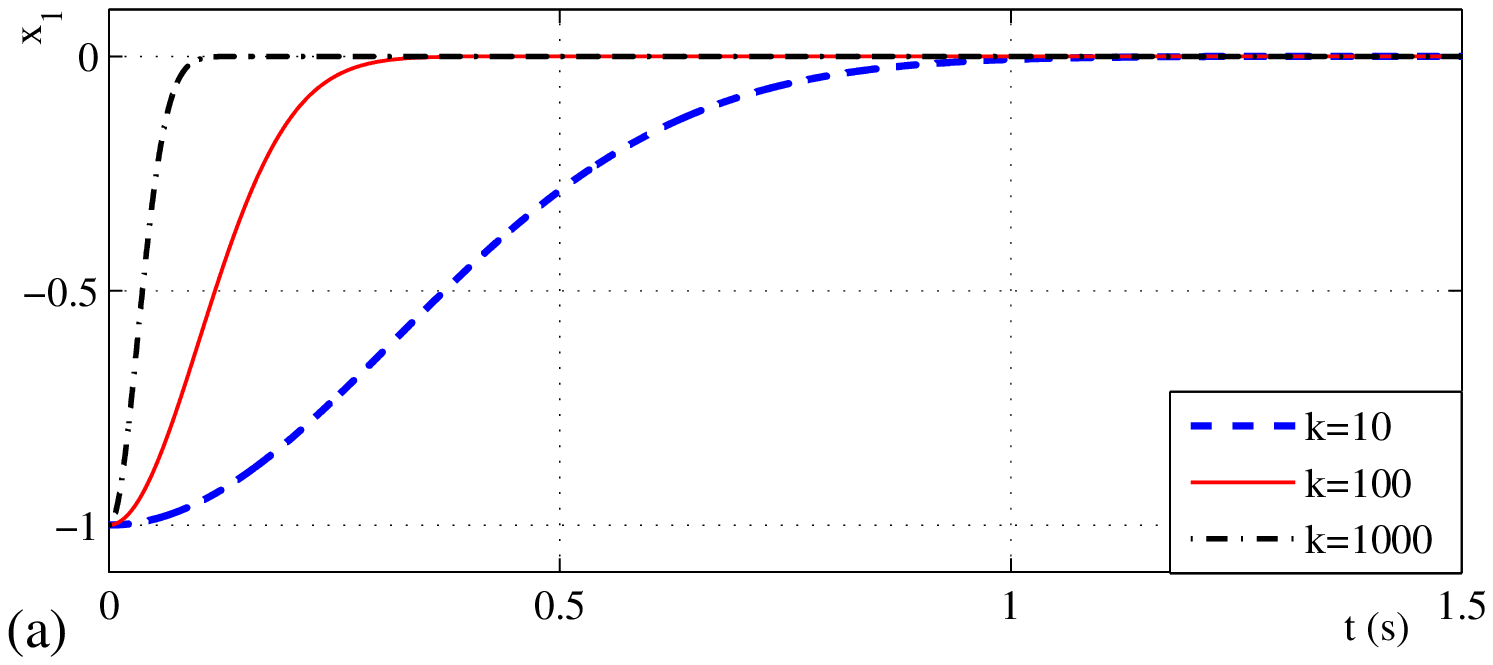}
\includegraphics[width=0.96\columnwidth]{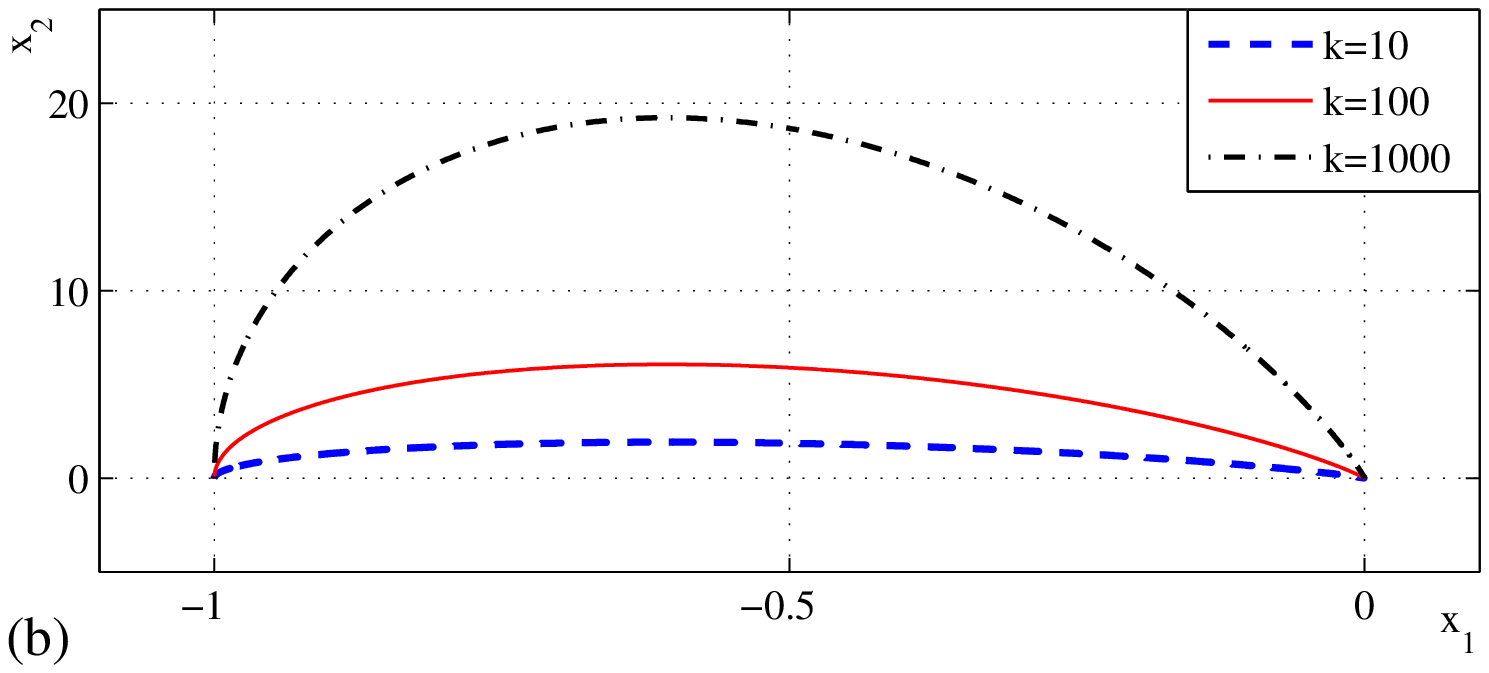}
\caption{Output convergence (a) and trajectories phase portrait
(b) of the nonlinear damping control with various $k$ parameter.}
\label{fig:6}
\end{figure}

The output convergence and state trajectories of the nonlinear
damping control are shown in Fig. \ref{fig:6} when assuming the
varying (by order) values of the feedback gain $k
=\{10,100,1000\}$. One can recognize a similar (scaled) trajectory
shape independent of the control gain value.

\begin{figure}[!h]
\centering
\includegraphics[width=0.98\columnwidth]{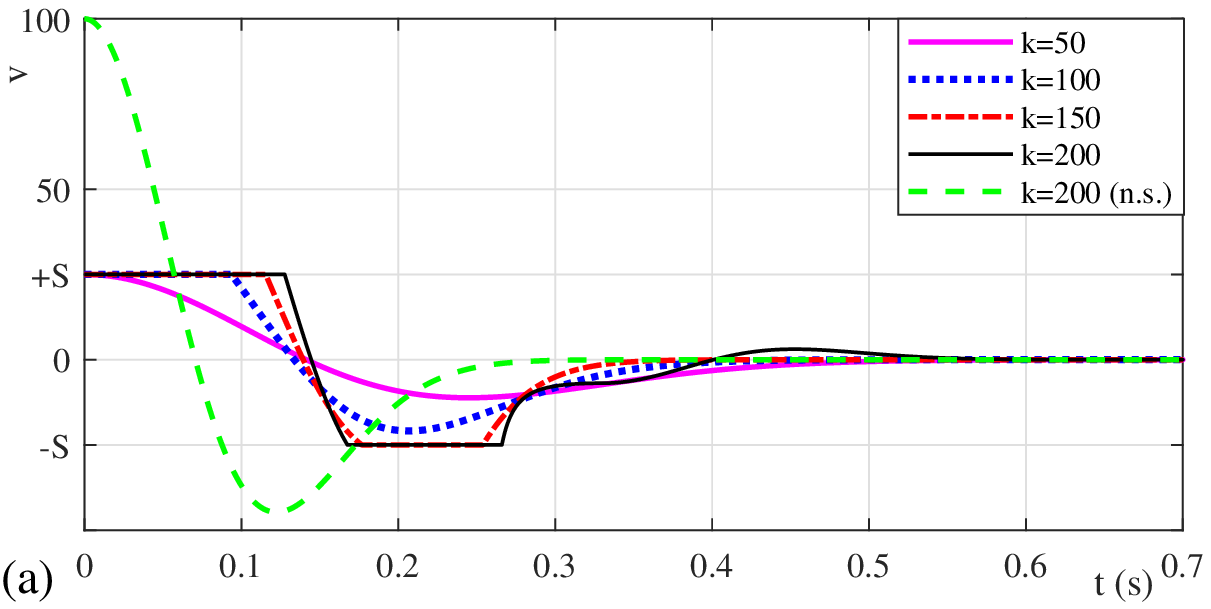}
\includegraphics[width=0.98\columnwidth]{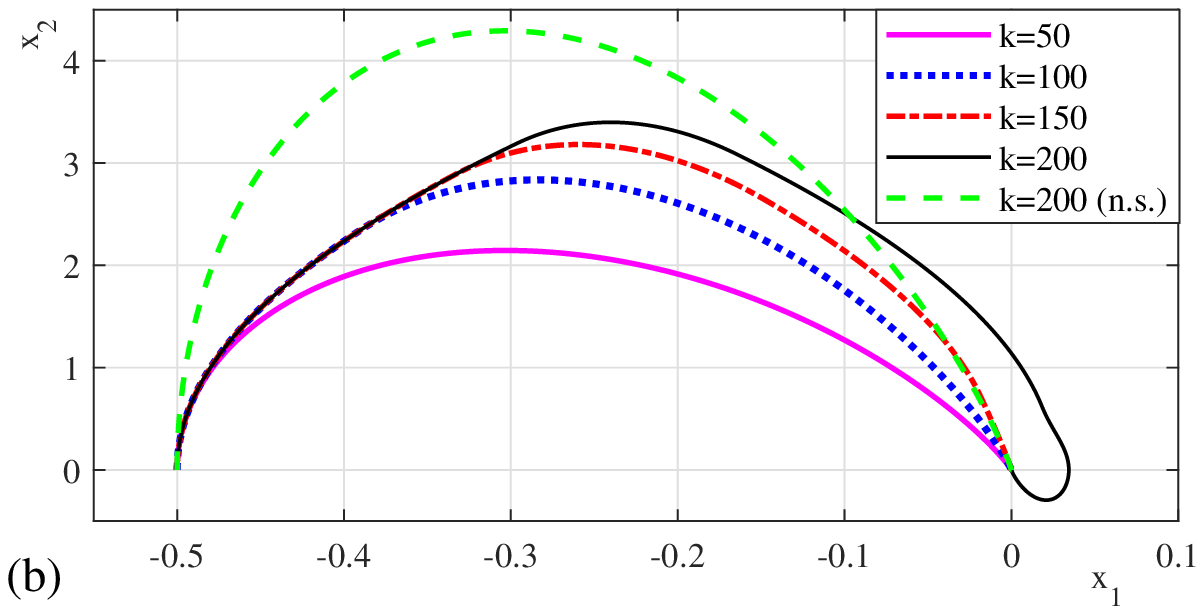}
\caption{Control value (a) and trajectories phase portrait (b) for
various $k$-parameter values, with and without control
saturation.} \label{fig:7}
\end{figure}

Finally, the impact of the control saturation, i.e. of the bounded
$v$-input cf. section \ref{sec:2:sub:5}, is demonstrated for the
different feedback gain values $k =\{50,100,150,200\}$ and $S=25$;
for the largest gain $k=200$ the non-saturated (n.s.) case, i.e.
$S=\infty$, is also included for the sake of comparison. The
control response and state trajectories are shown in Fig.
\ref{fig:7} (a) and (b) respectively. The saturation slows down
the convergence and leads, in worst case of largest gain, to a
single transient overshoot, after which the trajectory converges
as expected (cf. with Fig. \ref{fig:1}).

\section{Conclusions}
\label{sec:4}

This paper has proposed a novel nonlinear damping control for the
second-order unperturbed systems with output feedback. The control
is claimed to be optimal since it does not require any additional
parameter and provides a fast (exponentially quadratic)
convergence without transient overshoots, when no control
constraints. The global asymptotic stability, passivity, and
finite-time convergence until certain neighborhood to the stable
origin of the state variables have been explored. An enhanced
performance has been demonstrated comparing to the linear and
optimally (i.e. critically) damped controller. Also the saturated
control case, as relevant for applications, was analyzed,
regarding convergence, and demonstrated to have no negative impact
on the principal control performance.

\bibliographystyle{elsarticle-num}        

\bibliography{references}

\end{document}